\documentclass[10pt]{iopart}
\usepackage{graphicx}
\expandafter\let\csname equation*\endcsname\relax
\expandafter\let\csname endequation*\endcsname\relax
\usepackage{amsmath}
\usepackage{bm}
\usepackage{epsfig}
\usepackage{charter}
\usepackage{amssymb}
\usepackage{setspace}
\usepackage{csquotes}
\usepackage{float}
\usepackage{braket}
\usepackage{epstopdf}
\usepackage{xspace}

\begin{document}

\title{Observation of a neutron spin resonance in the  bilayered superconductor
  CsCa$_2$Fe$_4$As$_4$F$_2$}
\author{D.~T.~Adroja$^{1,2}$, S.~J.~Blundell$^3$, 
F.~Lang$^{1,3}$, H.~Luo$^{4,5}$, Z.-C.~Wang$^{6}$, G.-H.~Cao$^{6}$}

\address{$^1$ISIS Facility, STFC
Rutherford Appleton Laboratory,
Chilton,
Oxfordshire OX11 0QX,
United Kingdom}
\address{$^2$Highly Correlated Matter Research Group, Physics Department, University of Johannesburg, PO Box 524, Auckland Park 2006, South Africa}
\address{$^3$Oxford University Department of Physics, Clarendon Laboratory, Parks Road,
  Oxford OX1 3PU, United Kingdom} 
\address{$^4$Beijing National Laboratory for Condensed Matter Physics,
Institute of Physics, Chinese Academy of Sciences, Beijing 100190, China}  
\address{$^5$Songshan Lake Materials Laboratory, Dongguan, Guangdong 523808, China}
\address{$^6$Department of Physics and State Key Lab of Silicon
  Materials, Zhejiang University, Hangzhou 310027, China}

\begin{abstract}
  We report inelastic neutron scattering (INS) investigations on the  bilayer Fe-based superconductor CsCa$_2$Fe$_4$As$_4$F$_4$ above and below its superconducting transition temperature $T_{\rm c}\approx 28.9$~K to investigate the presence of a neutron spin resonance. This compound crystallises in a body-centred
  tetragonal lattice containing asymmetric double layers of
  Fe$_2$As$_2$ separated by insulating CaF$_2$ layers and is known to
  be highly anisotropic. Our  INS study clearly reveals the presence of a neutron spin resonance that exhibits higher intensity at lower momentum transfer ($Q$) at 5\,K compared to 54\,K,  at an energy of  15 meV.   The energy
  $E_{\rm R}$ of the observed spin resonance is broadly consistent with the
  relationship $E_{\rm R}=4.9 k_{\rm B} T_{\rm c}$, but is slightly enhanced compared to the values 
  observed in other Fe-based superconductors. We discuss the nature of the electron pairing symmetry by comparing the value of  $E_{\rm R}$ with that deduced from the total superconducting gap value integrated over the Fermi surface. \\
\end{abstract}
  ({\today})

\maketitle
\ioptwocol
%\tableofcontents

\section{Introduction}\label{Intro}
The discovery of iron-based superconductors in 2006 by Kamihara {\sl et al.}~\cite{Kamirara2006} has attracted considerable interest in condensed matter physics. Since then many families of iron-based superconductors have been reported \cite {MIZUGUCHI2010,Stewart2011,Dai2015, Bhattacharyya2019}. Despite the availability of many experimental results and various proposed theoretical models \cite {Mazin2008,Platt2012} the origin of the superconductivity and the nature of electron pairing symmetry in iron-based superconductors are still under debate.  It is widely believed that the interband interactions between the hole pockets at the zone center ($\Gamma$) and the electron pockets at the zone edges ($M$) play an important role in the electron pairing and
superconductivity in iron-based superconductors. The mechanism driving superconductivity in the iron-based materials is
currently thought to involve spin fluctuations which mediate the
electron pairing.  These spin fluctuations, mainly arising from the excitations between electron and hole pockets, may also give rise to a neutron
spin resonance centred around an energy ($E_{\rm R}$) which scales
linearly with the superconducting transition temperature $T_{\rm c}$
with the relationship $E_{\rm R} \sim 4.9 k_{\rm B}T_{\rm c}$
 in iron-based superconductors \cite {Dai2015,  Tao2018A}. This has been tested using a variety of iron-based
superconductors which fall into the well-known $1111$-, $111$-, $11$-, $112$- or
$122$-type families \cite {Dai2015, Christianson2008, Chi2009}. Furthermore, an inelastic neutron scattering study on CaKFe$_4$As$_4$ (1144-family) reveals the presence of three spin resonance modes and their intensity varies with the value of the momentum transfer (L) along the c-axis; these are hence called odd and even modes of the spin resonance \cite {Tao2018}. A very similar relation between  $E_{\rm R}$ and $T_{\rm c}$ has been observed in the cuprate high temperature superconductors (where the pairing symmetry is confirmed to be mainly d-wave pairing) with $E_{\rm R} \sim 5.8 k_{\rm B}T_{\rm c}$
 \cite{Zhao2007}. Furthermore,  heavy fermion superconductors, which have very low transition temperatures (below 3 K), also exhibit a neutron spin resonance \cite {Hap2016, Stock2015}.
 
The spin resonance mode is argued to be a collective mode originating from singlet-triplet excitations of the Cooper pairs, and thus the spin resonance can be considered to be a spin-exciton bound state at a resonant energy ($E_{\rm R}$) below the pair breaking energy (2$\Delta$) \cite{Tao2018A,Gyu2009,  Eschrig2006}.  Specifically in iron-based superconductors, the spin resonance mode was believed to be a key piece of evidence for the sign-reversal of the order parameter through Fermi surface nesting within the so-called weak coupling scenario \cite{ Mazin2008,Kuroki2008}. For this scenario, $E_{\rm R}$  should be within the total superconducting gap $\Delta_{\rm tot}$ which is obtained by summing over the nesting pockets linked by momentum $Q$: $\Delta_{\rm tot} = |\Delta_k| + |\Delta_k+Q|$.  Alternatively, a strong coupling approach, in which the pairing of electrons comes from short-range magnetic interactions, can give $s^{\pm}$-pairing and a spin resonance, while the gap function is completely different from the Fermi surface nesting picture \cite {Hirschfeld2011, Seo2008}. Although it is generally accepted that the resonance is a signature of unconventional superconductivity \cite {UCS}, there is no consensus on its microscopic origin. Moreover,  the spin resonance in iron-based superconductors has also been proposed to be a collective excitation with a magnon-like dispersion, which was later confirmed by neutron experiments \cite {magnon1, magnon2,magnon3,magnon4}. 

Very recently, a totally new structural
family was synthesized with general formula $A$Ca$_2$Fe$_4$As$_4$F$_2$, where $A=$~K,Rb,Cs.  They thus have what we can call a `$12442$
formula' and are superconductors with values of $T_{\rm c}$s up to 33~K 
\cite{Wang2016, Wang2017}.  The $A$Ca$_2$Fe$_4$As$_4$F$_2$ family of materials crystallise in a body-centred tetragonal lattice containing 
asymmetric double layers of Fe$_2$As$_2$ separated by insulating
CaF$_2$ layers. The alkali metal cations $A^+$ are sandwiched between
the Fe$_2$As$_2$ layers (see Fig.~1).
Note that the alkali-metal-containing ``122'' block is nominally hole
doped with 0.5 holes/Fe, while the ``1111'' block is nominally
undoped. Consequently, the 12442-type compounds are all hole self-doped
at a level of 0.25~holes/Fe.
This leads to superconductivity without the need for extrinsic doping;
in the case of extrinsic electron doping through Co/Fe substitution,
superconductivity gradually disappears, accompanied by a sign change in Hall coefficient \cite{hosono}.

By varying $A$, the structural parameters and atomic
coordinates of almost all of the atoms vary \cite{Wang2016, Wang2017},
and here we focus on $A$=Cs.  It was found that the superconductivity
is sensitive to these changes in the structural parameters, with
$T_{\rm c}$ inversely correlating with the lattice parameters $a$ and
$c$. This contrasts sharply with other hole-doped Fe$_2$As$_2$ systems
\cite{Chen,Luo} and so far the mechanism behind this variation is
unknown. As the effective intra-layer coupling increases with a
decrease in the lattice spacing, it is possible that
$T_{\rm c}$ is either enhanced by inter-bilayer coupling or suppressed
by intra-bilayer coupling, indicating that the superconducting
properties may not be solely dependent on single FeAs layers
\cite{Wang2017}. First principles electronic structure calculations of  CsCa$_2$Fe$_4$As$_4$F$_2$ reveal that the Fe-3d levels dominate the density of states at the Fermi level, and within these 3d states the contribution of e$_g$ states is significant, demonstrating the multi-band nature of this superconductor \cite {Singh2018}. The upper bound of the superconducting transition temperature, estimated using the electron-phonon  coupling constant, is found to be 2.6\,K. To produce the experimental value of transition temperature (28.9~K, see Section~2), a 4--5 times increase in the electron-phonon constant is necessary, hinting that conventional electron-phonon coupling is not enough to explain the origin of superconductivity in this material \cite{Singh2018}.

\begin{figure}[htb]
\centering
\includegraphics[width=0.7\columnwidth]{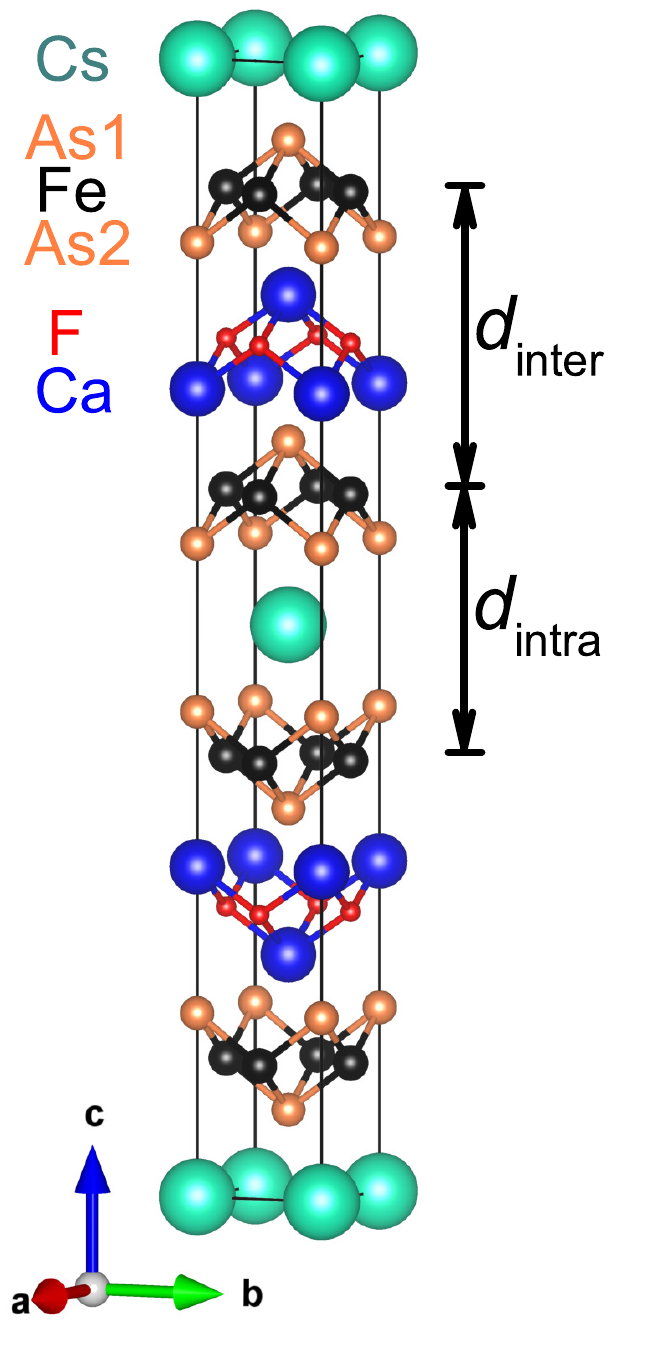}
\caption{The crystal structure of CsCa$_2$Fe$_4$As$_4$F$_2$ with separate double
Fe$_2$As$_2$ layers [After \cite{Wang2017}.)
}
\label{fig:wang}
\end{figure}

This paper reports the discovery of a spin resonance mode in a powder sample of CsCa$_2$Fe$_4$As$_4$F$_2$ using inelastic neutron scattering (INS).  
The compound contains the
largest alkali atom in this family of superconductors and it has been
characterised by muon-spin rotation which yielded a value for the
in-plane penetration depth of $\lambda_{ab}(0)=244(3)$~nm \cite{fkkk}.
The temperature evolution of the penetration depth strongly suggested
the presence of line nodes and is best modelled by a system consisting
of both an $s$- and a $d$-wave gap \cite{fkkk} (with similar results
for the Rb and K analogues \cite{smidman,adroja}), consistent with
multiband behaviour.  Recent single crystals measurements on CsCa$_2$Fe$_4$As$_4$F$_2$
have highlighted the giant anisotropy of the upper critical field H$_{c2}$ and the importance of
two-dimensional fluctuations in this material
\cite{Wang2019}.

\section{Experimental details}
A polycrystalline sample of CsCa$_2$Fe$_4$As$_4$F$_2$ was synthesized
via the solid-state reaction and details are given in Ref.~\cite{Wang2016}.  Our sample was found to
have a sharp superconducting transition at 28.9\,K, in agreement with the previous sample used in our $\mu$SR study
\cite{fkkk}.   
The INS measurements were carried out
using the high-neutron-flux MERLIN time-of-flight spectrometer at the
ISIS facility \cite{merlin}.  The powder sample (mass $\sim 4.1$~g) was
wrapped in a thin Al foil and mounted inside a thin-walled cylindrical
Al can with diameter of 30 mm (and height of 40 mm), which was cooled down in a closed cycle refrigerator (CCR) with He-exchange gas
around the sample. The measurements were performed with an incident neutron energy ($E_i$) 50~meV with chopper frequency of 350~Hz, which gave an elastic resolution of $\Delta$E =2.0~meV. The measurements were carried out at 5~K (below $T_c$) and at 54~K (above $T_c$) with counting time of 22 hours per temperature point.

\section {Results}
Figure~2(a,b) shows 2D colour maps of the inelastic neutron scattering response,  plotted as momentum transfer versus energy transfer, from the CsCa$_2$Fe$_4$As$_4$F$_2$ powder sample measured at below $T_{\rm c}$, 5 K and above $T_{\rm c}$, 54 K.
It is clearly seen that there is more scattering at low-$Q$ near 15~meV in the 5~K data, than that of 50 K data (see Fig.3 for 1D cuts). We attribute the excess scattering near 15 meV at 5~K compared with 54~K to the presence of the spin resonance mode below $T_{\rm c}$. Note that the intensity of the spin resonance mode in CsCa$_2$Fe$_4$As$_4$F$_2$ seems weaker compare to that observed in Ba$_{0.6}$K$_{0.4}$Fe$_2$As$_2$ \cite {Christianson2008} and CaKFe$_4$As$_4$ \cite{Iida2017, Tao2018}.
In order to see clearly the presence of the spin resonance and excess magnetic scattering intensity at 5~K compared with 54~K, we have taken the temperature difference of the two data sets, i.e. 5 K minus 54 K.  Figure~4a shows the 2D colour maps of the difference, 5 K$-$54 K, inelastic neutron scattering intensity, which clearly reveals the presence of magnetic scattering near 15~meV.  The magnetic nature of the scattering can be deduced by noting the fall in the intensity of this feature with increasing momentum transfer. The energy-integrated (between 13 and 17 meV) one dimensional momentum ($Q$) cut showed a peak at Q=1.35(2)~\AA$^{-1}$ [see Fig.4(b)]. 
%Fig.~\ref{fig:s2}%
 
 Fig.~4(c) presents an energy cut through the temperature difference data (5 K minus 54 K) from low scattering angles, integrated over 7--20 degrees, that again reveals the presence of a clear sign of the  spin resonance at $\sim15$~meV. We note that the cut was made in scattering angle rather than $Q$ to avoid the spurious scattering from the first few detector tubes at low scattering angles. It is clear from Fig.~4(c) that the integrated scattering intensity between 12 meV and 20 meV shows positive values. It is very difficult from our data to determine whether there is more than one resonance peak present in CsCa$_2$Fe$_4$As$_4$F$_2$. Considering the double layer structure of CsCa$_2$Fe$_4$As$_4$F$_2$, we would expect more than one resonance peak in the 12442-family due to multiple Fe d-bands contributing at the  Fermi level. This has been supported through first-principle calculations on electronic structure and magnetic properties of KCa$_2$Fe$_4$As$_4$F$_2$ \cite{GuangtaoWang2016}, which reveal a more complicated  FS than other FeAs-based superconductors. Here there are ten bands crossing the Fermi level in the nonmagnetic state, resulting in six hole-like FS sheets along $\Gamma-Z$ and four electron-like sheets along $X-P$. The fixed spin moment calculations and the comparisons between total energies of different magnetic phases indicate that KCa$_2$Fe$_4$As$_4$F$_2$ has a strong tendency towards a stripe antiferromagnetic state. It has been found that the self-hole-doping suppresses the spin-density wave (SDW) state, inducing superconductivity in the parent compound KCa$_2$Fe$_4$As$_4$F$_2$~\cite{GuangtaoWang2016}.  Furthermore, a comprehensive angle-resolved photoemission spectroscopy (ARPES) study on KCa$_2$Fe$_4$As$_4$F$_2$ shows divergent superconducting gaps with splitting electron bands on different sized Fermi pockets \cite{Wu-arXiv}. 
 
 It is interesting to note that three resonance peaks were observed in the inelastic neutron scattering study on single crystals of CaKFe$_4$As$_4$ due to the very different superconducting gaps on the nesting pockets \cite{Tao2018}. Furthermore, a powder inelastic neutron scattering study of CaKFe$_4$As$_4$ \cite{Iida2017} also reveals a double peak type structure, near $\sim10$~meV and $\sim15$~meV,  with magnetic intensity extending up to $\sim25$~meV. The presence of multiple spin resonances has been also supported by a random phase approximation (RPA) calculation on CaKFe$_4$As$_4$ that reproduces the broad spectral features of the spin resonance anticipated as a result of multiple superconducting gaps on the different Fermi surface sheets in CaKFe$_4$As$_4$ \cite{Iida2017}. It is observed for all the Fe-based compounds that the excess intensity due to the spin resonance mode disappears at $T_{\rm c}$. A more detailed study of the temperature dependence of the spin resonance in our material would be an interesting subject for a future study which would be most informative if it could be carried out on single crystal samples.
 
 We note that the resonance peak we observe at $Q=1.35(2)$\,\AA$^{-1}$ is larger than the wave vector from the $\Gamma$ to the M point ($|{\bf Q}|=|(0.5, 0.5, 0)|\,\mbox{r.l.u.}= 1.15$\,\AA$^{-1}$), which  is usually the zone center of magnetic fluctuations in 122 and 1144-type iron-based superconductors.  However, if we consider an incommensurate resonance peak along the H direction, as is found in single crystal results on KCa$_2$Fe$_4$As$_4$F$_2$ \cite{HLuo2020}, a reasonable explanation can be given. For example, assuming incommensurate resonance peaks at ${\bf Q}=(0.4, 0.4, 0)$ and ${\bf Q}=(0.6, 0.6, 0)$, respectively, as we have fewer detectors at the lower-$Q$ side for $E=$13--17\,meV (data from bad detectors have been removed), it is possible to miss lower $Q$-scattering at ${\bf Q}=(0.4, 0.4, 0)$ (about 0.92\,\AA$^{-1}$) and mainly measure the  higher-$Q$ peak at ${\bf Q}=(0.6, 0.6, 0)$ (about 1.37\,\AA$^{-1}$).
 
\begin{figure} 
\centering
\includegraphics[width=1.0\columnwidth, trim={25mm 70mm 25mm 30mm},clip]{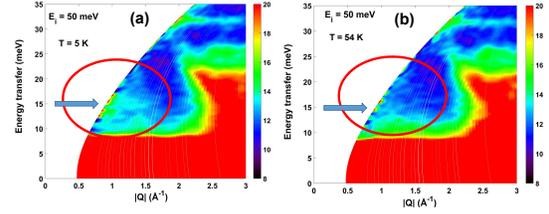}
\caption{Inelastic scattering data at (a) 5\,K and (b) 50\,K.  There
  is more scattering near 15~meV at 5\,K than at 50\,K, shown by arrow and circle, indicating
  the presence of a spin resonance.  The incident neutron energy is 50\,meV.
}
\label{fig:Fig2}
\end{figure}

\begin{figure}
\centering
\includegraphics[width=1.0\columnwidth]{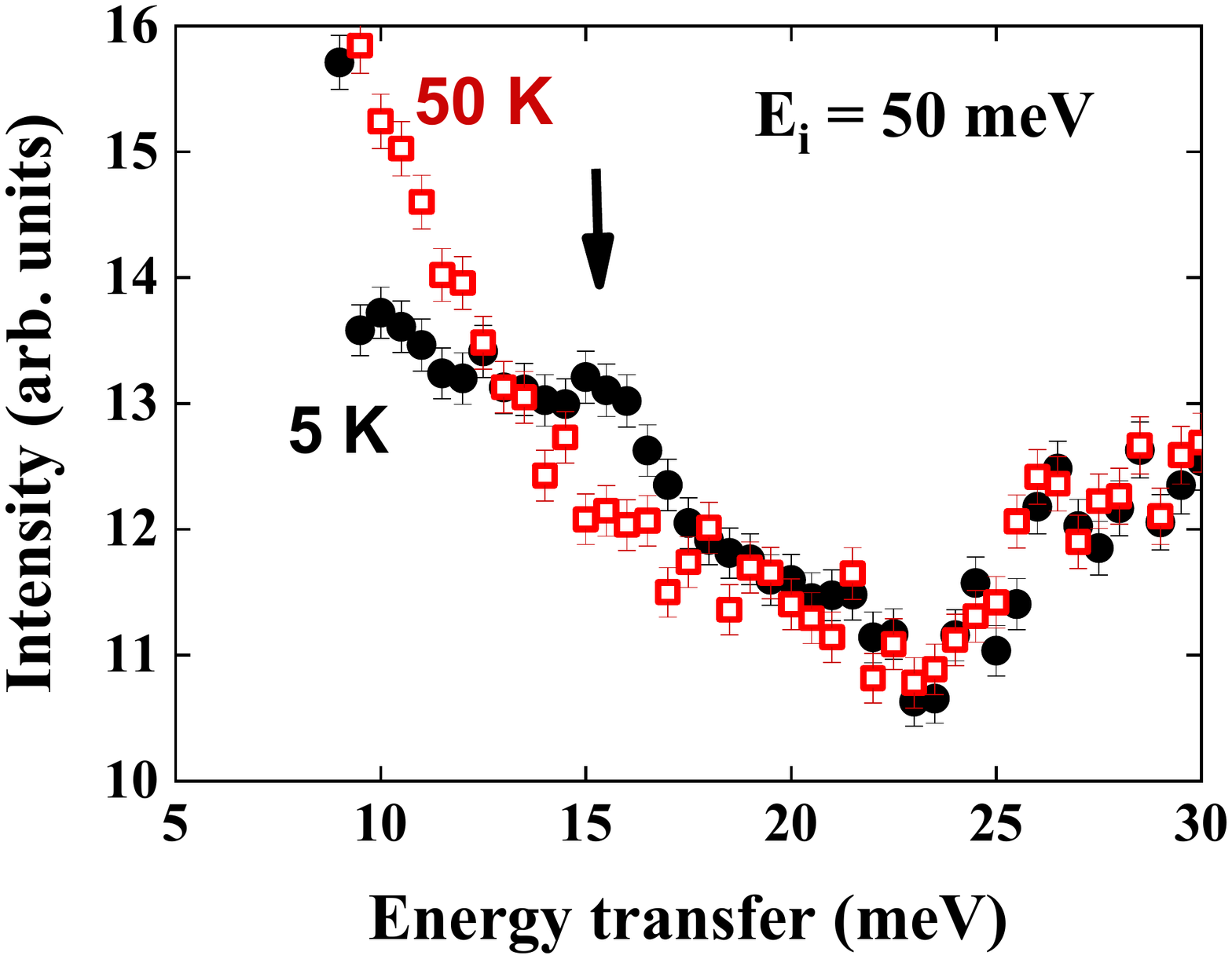}
\caption{One dimensional cuts taken from the data presented in Fig.2 for scattering
  angles between 7 and 20 degrees, to avoid bad detectors at low angles.  The presence of the spin resonance at $\sim 15$~meV is clear in the 5\,K data.  The data are from incident neutron energy of 50\,meV.
}
\label{fig:Fig3}
\end{figure}

\begin{figure}
\centering
\includegraphics[width=1.0\columnwidth]{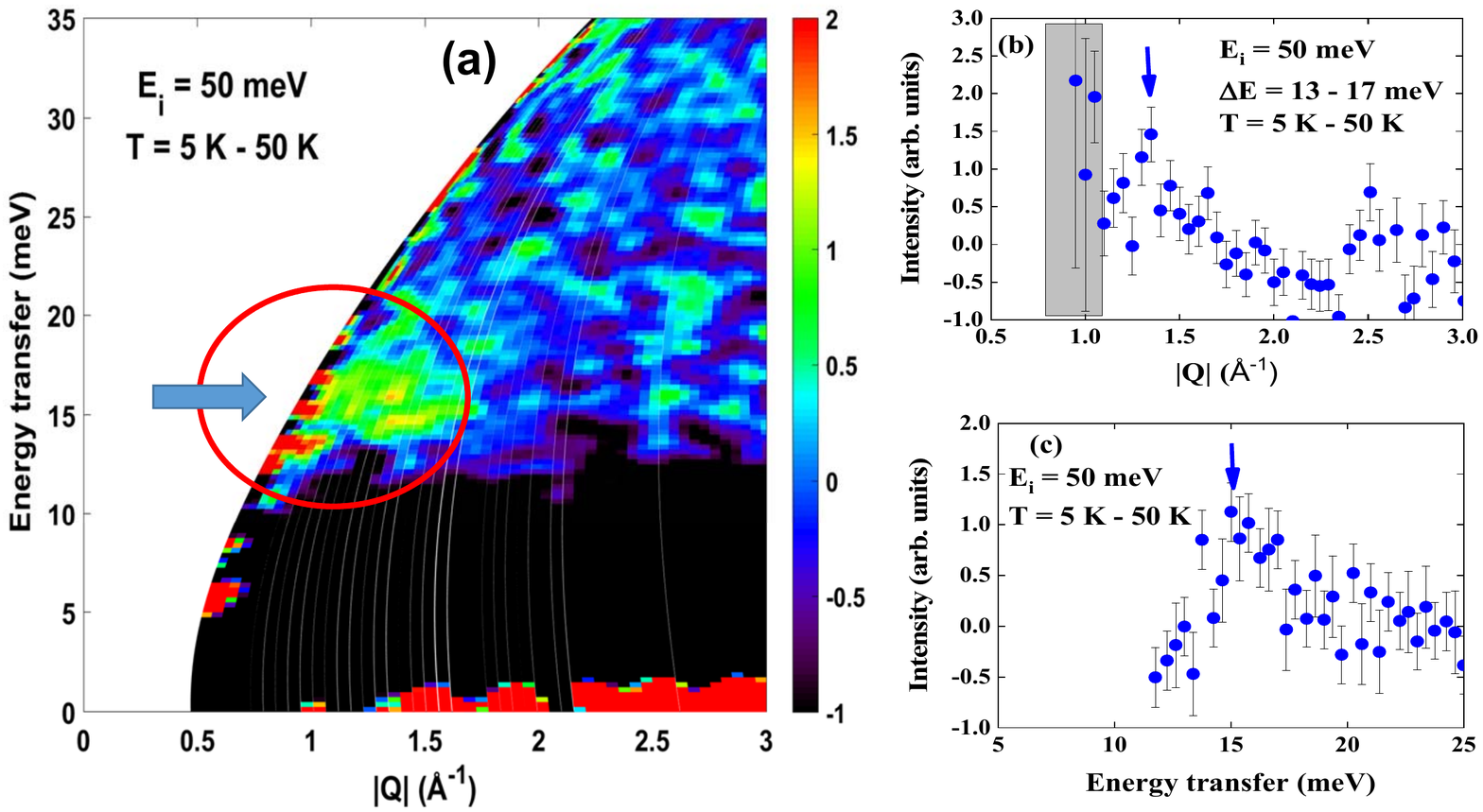}
\caption{(a) The temperature difference colour map of the scattering intensity, between the 5\,K  and 50\,K data  is
  plotted as energy transfer vs momentum transfer. The arrow and circle indicate the presence of spin resonance and, (b) One dimensional Q-cut taken from the data presented in (a) energy integrated between 13 and 17 meV. The shaded region shows the data from the bad detectors at low-Q (or low-angles).(c) One dimensional Energy-cut taken from the data presented in (a) for scattering
  angles between 7 and 20 degrees, to avoid bad detectors at low angles. The presence of the neutron spin resonance at $\sim 15$~meV is clear in the temperature difference data.
}
\label{fig:s3}
\end{figure}

\section{Discussion}
The observation of a neutron spin resonance is usually taken as  crucial evidence for spin-fluctuation mediated superconducting pairing in unconventional superconductors \cite {Dai2015}. The spin resonance mode is argued to be a collective mode from singlet-triplet excitations of the Cooper pairs, so that the spin resonance can be consider as a spin-exciton bound state at resonant energy ($E_{\rm R}$) below the pair breaking energy (2$\Delta$) \cite{Gyu2009, Tao2018A, Eschrig2006}.  For $s^\pm$ pairing symmetry the spin resonance energy $E_{\rm R} < 2\Delta(k,Q) = |\Delta_k| + |\Delta_{k+Q}|$, where the gap function is $\Delta_k=\Delta_0\cos k_x \cos k_y$ or $\Delta_k=\Delta_0(\cos k_x+\cos k_y)/2$. On the other hand for $s^{++}$ pairing symmetry $E_{\rm R} > 2\Delta(k,Q)$, so that a resonance-like peak can emerge above $2\Delta(k,Q)$ (maximum is $2\Delta_0$). The superconducting gaps estimated from the $\mu$SR for s+d  wave pairing in this compound are $\Delta$$^s$=7.5 meV and $\Delta$$^d$=1.5 meV, which would imply $E_{\rm R} < 2\Delta (k,Q)$, which may challenge the picture of $s^\pm$ pairing symmetry in CsCa$_2$Fe$_4$As$_4$F$_2$. Further investigation on the gap function in momentum space is highly desirable to clarify this issue.

As has been observed for many Fe-based superconductors the value of neutron spin resonance energy $E_{\rm R}$  is directly related to $T_{\rm c}$. For CsCa$_2$Fe$_4$As$_4$F$_2$ we have $E_{\rm R}/(k_{\rm B}T_{\rm c}) = 6.02$, 
which is slightly higher than the value of 4.9 observed for other iron-based superconductors  \cite{johnson2015}.
Recently the neutron spin resonance has also been investigated in KCa$_2$Fe$_4$As$_4$F$_2$  single crystals with $T_{\rm c}$=33.5 K \cite{HLuo2020}, which reveals a spin resonance $E_{\rm R}$=16 meV. This give the ratio $E_{\rm R}/(k_{\rm B}T_{\rm c}) =5.5$ for KCa$_2$Fe$_4$As$_4$F$_2$ in broad agreement with what we have observed  for CsCa$_2$Fe$_4$As$_4$F$_2$.  The neutron spin resonance found in CsCa$_2$Fe$_4$As$_4$F$_2$ can be
plotted on a graph of $E_{\rm R}$ against $T_{\rm c}$ (see
Fig.~5) and compared with data obtained on a variety of other iron-based superconductors
(collected in \cite{Tao2018}). The new data point falls close to the
correlation line $E_{\rm R}=4.9 k_{\rm B} T_{\rm c}$, but with a small enhancement, possibly associated with the complex gap function in terms of a splitting on specific bands and strong bilayer couplings \cite{Wu-arXiv} (which would apply also to KCa$_2$Fe$_4$As$_4$F$_2$).  We note that the correlation found for d-wave pairing symmetry in the cuprates is $E_{\rm R}=5.8 k_{\rm B} T_{\rm c}$, so that this enhancement may be connected with the s+d-pairing symmetry in this iron-based compound, which requires further studies on the momentum-resolved superconducting gaps.  An alternative explanation arises from the highly two-dimensional nature of this superconductor.  The resulting large anisotropy may be associated with weak coupling along the $c$-axis, which could reduce the bulk $T_{\rm c}$ value without similarly affecting $E_{\rm R}$, thereby slightly enhancing the value of $E_{\rm R}/k_{\rm B}T_{\rm c}$.

\begin{figure}[htb]
\centering
\includegraphics[width=1.0\columnwidth]{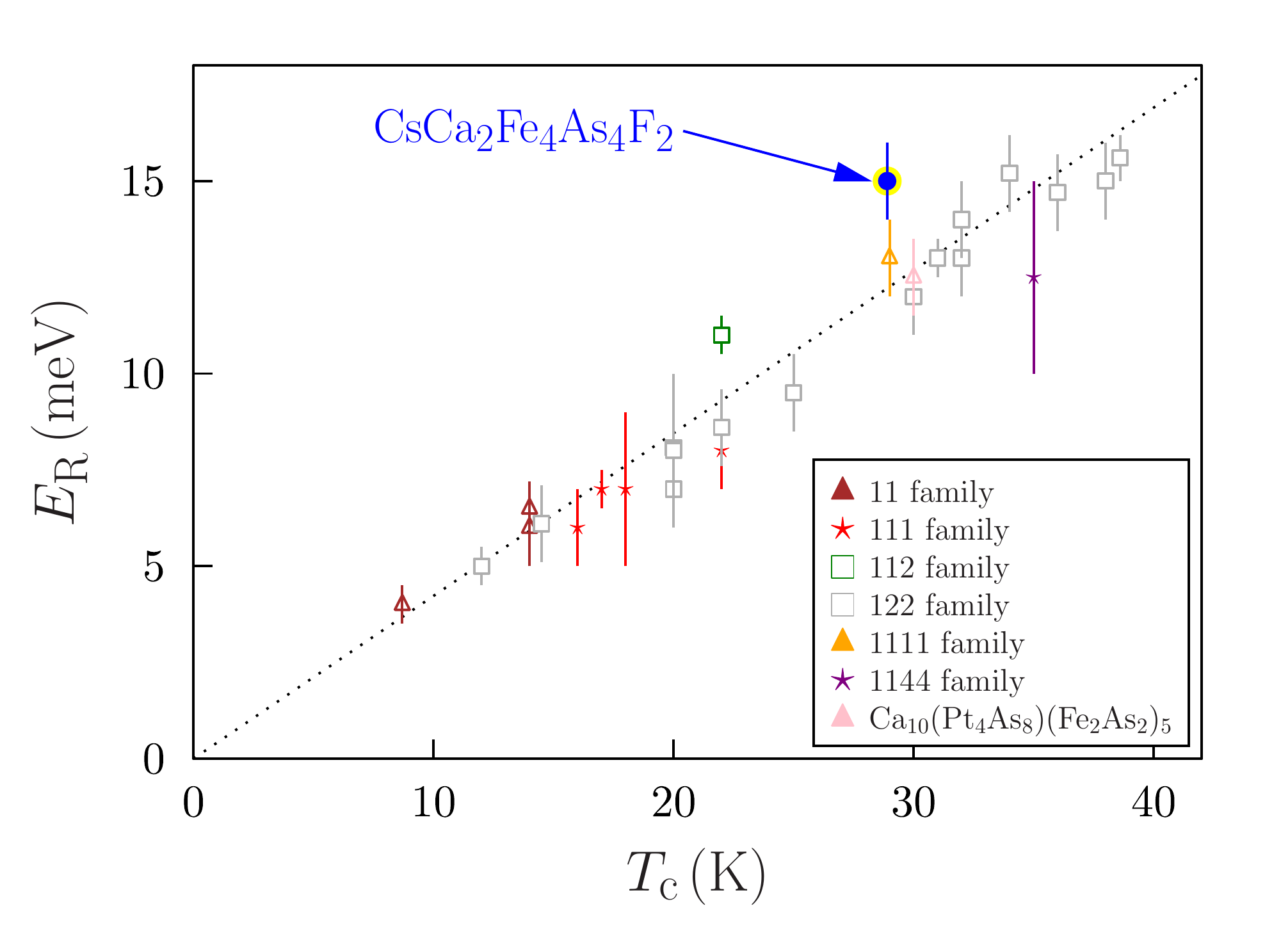}
\caption{The neutron spin resonance energy $E_{\rm R}$ as a function of $T_{\rm c}$ for a collection of iron-based superconductors collated from \cite{Tao2018A,johnson2015}, together with the new data point from the present work. The straight line assumes the relationship $E_{\rm R}=4.9 k_{\rm B} T_{\rm c}$.
}
\label{fig:Fig_5}
\end{figure}

\section{Conclusion}        
We have reported inelastic neutron scattering (INS) results on the  bilayer Fe-based superconductor CsCa$_2$Fe$_4$As$_4$F$_4$ above and below its superconducting transition temperature $T_{\rm c}\approx 28.9$~K to investigate the presence of the neutron spin resonance. Our  INS study clearly reveals the presence of a neutron spin resonance, showing higher intensity at lower momentum transfer ($Q$) at 5 K compared to 54 K,  at energy of  15 meV.   The energy $E_{\rm R}$ of the observed spin resonance is broadly consistent with the relationship $E_{\rm R}=4.9 k_{\rm B} T_{\rm c}$ that has been observed in other Fe-based superconductors, but our value is slightly enhanced, consistent with an effect observed in a KCa$_2$Fe$_4$As$_4$F$_4$ single crystal sample. This enhancement needs further investigation but we suggest that it may be an effect associated either with the nature of the pairing found in this structural type or with the high two-dimensionality.

\section{Acknowledgments}
We thank the EPSRC (UK) for financial support under grant EP/N023803/1 and the ISIS Facility for beam time on MERLIN, RB1820214. The experimental data can be obtained at  DOI:10.5286/ISIS.E.RB1820214.  D.T.A. and H.L. would like to thank  the Royal Society of London for the UK-China Newton mobility funding.

\section*{References}

\end{document}